\documentclass[aps,pre,twocolumn,showpacs,superscriptaddress,floatfix]{revtex4}
\usepackage{amsmath}
\usepackage{amsfonts}
\usepackage{graphicx}

\begin{document}
%\svnInfo $Id: paper.tex 92 2009-05-05 15:35:51Z gert $ 

\title{Quantum dissipative Brownian motion and the Casimir effect}
\author{Gert-Ludwig Ingold}
\affiliation{Institut f\"ur Physik, Universit\"at Augsburg, D-86135 Augsburg}
\affiliation{Laboratoire Kastler Brossel, CNRS, ENS, UPMC, Campus Jussieu Case 74,
F-75252 Paris Cedex 05, France}
\author{Astrid Lambrecht}
\author{Serge Reynaud}
\affiliation{Laboratoire Kastler Brossel, CNRS, ENS, UPMC, Campus Jussieu Case 74,
F-75252 Paris Cedex 05, France}
\begin{abstract}
We explore an analogy between the thermodynamics of a free dissipative quantum
particle and that of an electromagnetic field between two mirrors of finite
conductivity. While a free particle isolated from its environment will
effectively be in the high-temperature limit for any nonvanishing temperature,
a finite coupling to the environment leads to quantum effects ensuring the
correct low-temperature behavior. Even then, it is found that under appropriate
circumstances the entropy can be a nonmonotonic function of the temperature.
Such a scenario with its specific dependence on the ratio of temperature and
damping constant also appears for the transverse electric mode in the Casimir
effect. The limits of vanishing dissipation for the quantum particle 
and of infinite conductivity of the mirrors in the Casimir effect both turn out 
to be noncontinuous. 
\end{abstract}

\maketitle

\section{Introduction}
Recently there has been considerable interest in the study of thermodynamic
quantities of a system in the presence of a finite coupling to the heat bath
\cite{hangg06,hoerh08,bandy08,wang08,hangg08,kumar09,ingol09,bandy09}. In
contrast to the classical equilibrium state, the stationary quantum state
depends on the system-bath coupling and thus the low-temperature regime is of
particular interest. Even on a conceptual level, for finite system-bath
coupling, the correct definition of basic thermodynamic quantities like the
specific heat has proven to be far from obvious \cite{hangg06,hangg08}.

The thermodynamics of a free particle coupled to a heat bath is particularly
rich.  In the absence of an environment, the free particle behaves classically
for any nonvanishing temperature. This unusual behavior is a consequence of the
lack of any energy scale besides the thermal energy $k_BT=1/\beta$ where $T$ is
the temperature and $k_B$ is the Boltzmann constant. The situation changes once
the particle is confined to a finite region in space of length $L$ or coupled
to an environment. In the first case, an energy scale $\hbar^2/2ML^2$ appears
where $M$ is the mass of the particle, while in the latter case, the relaxation
frequency $\gamma$ due to the coupling to the environment yields an energy
scale $\hbar\gamma$. 

It is the second scenario which one is interested in when studying the effect
of a finite coupling to the heat bath. Then all thermodynamic quantities will
depend on the dimensionless temperature $k_BT/\hbar\gamma$. As a first example
of the intricacy associated with the limit of vanishing damping, $\gamma\to0$,
in such a situation, we mention the diffusion of a free Brownian particle. The
long-time behavior is governed by the diffusion constant $D=k_BT/M\gamma$. For
any nonvanishing value of $\gamma$, one finds diffusive motion while for
$\gamma=0$, the motion will be ballistic. The limit $\gamma\to0$ will thus be
qualitatively different from $\gamma=0$.

The peculiar combination $k_BT/\hbar\gamma$ implies that for any nonzero
temperature, the free particle will be driven into the classical regime when
$\gamma$ approaches zero. On the other hand, for any nonzero value of $\gamma$,
there exists a low-temperature region which is essentially quantum in nature
and depends on the coupling to the heat bath \cite{hangg06,hangg08}. In this
regime, the specific heat approaches zero as temperature goes to zero while the
specific heat of an isolated free particle remains at its classical value
$k_B/2$. 

Interestingly, for appropriately chosen coupling between the free particle and
the heat bath, the entropy exhibits a nonmonotonic behavior as a function of
temperature. Evaluation of the specific heat according to standard
thermodynamic relations results in negative values. However, as entropy and
specific heat pertain only to a subsystem formed by the free particle, this is
not in contradiction to basic thermodynamic stability criteria \cite{ingol09}.

While the free particle might appear to be rather special, we will argue in the
present paper that the scenario just discussed also applies in the context of
the Casimir effect (for a review see \cite{borda01,milto04,njp06,lambr06} and
references therein). There, in the simplest case, one considers the force
between two parallel plane mirrors separated by a distance $L$ due to the
boundary conditions imposed by the mirrors on the electromagnetic field
\cite{casim48}. The coupling between the electrons in the mirrors and the
electromagnetic field therefore is essential, in particular if real mirrors are
considered. On the one hand, the reflectivity of metals goes to zero for high
frequencies on a scale determined by the plasma frequency $\omega_P$. On the
other hand, as long as normal metals are considered, the dc conductivity
$\sigma_0$ is finite. Within the Drude model this finite conductivity
unavoidably comes with a relaxation frequency $\gamma$.

The conductivity of metals like gold employed in measurements of the Casimir
force is so large that the relaxation frequency $\gamma$ is much smaller than
the plasma frequency $\omega_P$ and smaller than the frequency $c/L$ associated
with mirror separations between 20\,nm and $6\,\mu\text{m}$ used in
experiments so far \cite{decca07,munda07,chan08,vanzw08,munda09,jourd09,masud09,deman09}. 
On the other hand, at room temperature the energy scales $k_BT$ and $\hbar\gamma$ 
are comparable, thus opening the opportunity to explore the transition from the 
classical to the quantum regime.

As we will show in this paper, if apart from the thermal scale the relaxation
frequency $\gamma$ represents the smallest frequency scale, essential
low-temperature features of the Casimir effect within the Drude model are
analogous to those pointed out above for the free Brownian particle.  In our
opinion, this analogy will be valuable for the understanding of the thermal
Casimir effect and shed additional light on thermodynamic peculiarities related
to the transverse electric mode \cite{klimc06,brevi06,milto08,torge04} like the 
negative entropy found for the Drude model. 

We will start in Sec.~\ref{sec:freeparticle} by reviewing recent results on the
thermodynamics of the free Brownian particle and presenting new results on the
entropy which is of particular interest in relation to the Casimir effect. In
Sec.~\ref{sec:drudemodel}, essential properties of a Drude metal like its
permittivity and reflectivity are introduced. Sec.~\ref{sec:thermalcorr} is
devoted to the thermal corrections to the zero-temperature Casimir force. The
low-temperature behavior of the Casimir force is discussed and the leading
thermal correction for the transverse electric mode within the Drude model is
obtained on the basis of a scattering approach.  In addition, the leading
thermal correction for the transverse magnetic mode is presented. In
Sec.~\ref{sec:temode}, we establish the analogy between the free Brownian
particle and the thermal corrections to the Casimir force due to the transverse
electric mode within the Drude model. This analogy will be further discussed in
Sec.~\ref{sec:entropy} in terms of the free energy and the entropy. Finally,
in Sec.~\ref{sec:conclusions} we present our conclusions. Technical details
concerning the derivation of the leading thermal corrections to the Casimir
force within the Drude model will be given in an appendix.

\section{Thermodynamics of the free Brownian particle}
\label{sec:freeparticle}

As a paradigm for the peculiar effects of a finite system-bath coupling on the
thermodynamics of the system degree of freedom, we consider a free particle of
mass $M$. For normalization purposes, the particle will be confined to
a region of size $L$ which is assumed to be sufficiently large so that the
energy quantization is irrelevant for the temperatures of interest,
\textit{i.e.}\ $\hbar^2\beta/2mL^2\ll1$. Even for microscopic systems like a
hydrogen atom and the lowest temperatures attainable today, choosing
$L=1\,\text{cm}$ would easily meet this condition.

The free particle is assumed to be bilinearly coupled to a bath of harmonic oscillators.
The total Hamiltonian can then be cast into the form
\begin{equation}
H = H_S+H_B+H_{SB}
\end{equation}
with the system Hamiltonian
\begin{equation}
\label{eq:hs}
H_S = \frac{P^2}{2M}
\end{equation}
where $P$ is the momentum of the particle, the bath Hamiltonian describing an infinite 
collection of harmonic oscillators of masses $m_n$ and frequencies
$\omega_n$, 
\begin{equation}
\label{eq:hb}
H_B = \sum_{n=1}^{\infty}\frac{p_n^2}{2m_n}+\frac{m_n}{2}\omega_n^2x_n^2\,,
\end{equation}
and the Hamiltonian 
\begin{equation}
\label{eq:hsb}
H_{SB}=\sum_{n=1}^\infty -m_n\omega_n^2x_nQ+\frac{m_n\omega_n^2}{2}Q^2
\end{equation}
coupling the free particle and the bath oscillators via their respective
positions $Q$ and $x_n$.  For a more detailed discussion of this Hamiltonian,
we refer the reader to the literature \cite{dittr98,weiss99,ingol02}. It is
worth noting that the second term appearing in (\ref{eq:hsb}) is crucial
to ensure the translational invariance of the free Brownian particle.
Furthermore, the bath parameters $m_n$ and $\omega_n$ are sufficient to model
any linear heat bath \cite{grabe88}.

As a starting point for the following thermodynamic considerations, we introduce
the reduced partition function \cite{hangg06,grabe88,calde83,grabe84,ford85}
\begin{equation}
\label{eq:partfunc}
\mathcal{Z} = \frac{\text{Tr}_{S+B}[\exp(-\beta H)]}{\text{Tr}_B[\exp(-\beta
H_B]}\,.
\end{equation}
In the absence of any coupling between system and bath, this quantity reduces
to the partition function of the system while otherwise it accounts for the
influence of the coupling. Any quantity which can be obtained by a linear
operation from the logarithm of this partition function can be viewed as a
difference of the quantities related to the system and bath on one hand and to
the bath alone on the other hand \cite{ingol09}. It thus describes the effect
induced by coupling a system degree of freedom to the heat bath. Such
quantities therefore do not need to satisfy thermodynamic conditions which have
to be imposed on closed systems.

For a free Brownian particle, the reduced partition function defined in
(\ref{eq:partfunc}) is given by \cite{hangg08}
\begin{equation}
\label{eq:partfuncfree}
\mathcal{Z} = \frac{L}{\hbar}\left(\frac{2\pi m}{\beta}\right)^{1/2}
\prod_{n=1}^\infty\frac{\xi_n}{\xi_n+\hat\gamma(\xi_n)}\,.
\end{equation}
Here,
\begin{equation} 
\label{eq:matsubarafreq}
\xi_n=\frac{2\pi n}{\hbar\beta}
\end{equation}
are the Matsubara frequencies and $\hat\gamma$ is the Laplace transform of the
damping kernel which for ohmic damping is simply given by a constant,
\textit{i.e.}\ $\hat\gamma(\xi)=\gamma$. In order to ensure the convergence of
the infinite product, a high-frequency cutoff has to be introduced which is
often chosen to be of the form \cite{footnote1}
\begin{equation}
\label{eq:drudekernel}
\hat\gamma(\xi) = \frac{\gamma\omega_c}{\xi+\omega_c}\,.
\end{equation}
In the internal energy and therefore also the free energy, a finite cutoff
$\omega_c$ is needed while the limit of infinite cutoff can be taken for the
entropy, for example. As we will see at the end of this section, it may however
be interesting to keep $\omega_c$ finite and even to consider the case of very
small cutoff frequencies.

Based on (\ref{eq:partfuncfree}), we define an entropy $S$ by means of the standard
thermodynamic relation
\begin{equation}
\label{eq:s_def}
\frac{S}{k_B} = \frac{\partial}{\partial T}[k_B T\log(\mathcal{Z})]\,.
\end{equation}
We will carry out the discussion of the entropy in two steps which both lead to 
results of relevance for our later considerations of the Casimir effect. First, we 
will restrict ourselves to the limit of infinite cutoff frequency $\omega_c$ and then,
in a second step, we admit finite cutoff frequencies.

From (\ref{eq:s_def}) one finds with (\ref{eq:partfuncfree}) and (\ref{eq:drudekernel}) in
the limit of $\omega_c\to\infty$ the entropy
\begin{equation}
\label{eq:s_fp}
\begin{aligned}
\frac{S}{k_B} &= \frac{S_0}{k_B}+\log\left[\Gamma\left(1+\frac{\hbar\beta\gamma}{2\pi}
\right)\right]-\frac{\hbar\beta\gamma}{2\pi}\psi\left(\frac{\hbar\beta\gamma}{2\pi}\right)\\
&\quad-\frac{1}{2}\log(\hbar\beta\gamma)+\frac{\hbar\beta\gamma}{2\pi}-\frac{1}{2}\,,
\end{aligned}
\end{equation}
where $\Gamma(x)$ is the gamma function and $\psi(x)$ its logarithmic derivative.
In the zero-temperature limit, the entropy takes the value
\begin{equation}
\label{eq:s0}
\frac{S_0}{k_B} = \frac{1}{2}\log\left[2\pi\frac{ML^2\gamma}{\hbar}\right]
\end{equation}
which depends logarithmically on the box size $L$. The existence of this non-vanishing 
entropy $S_0$ is a consequence of the fact that we do not account for the level spacing 
due to the finite box size as discussed at the beginning of this section. We rather are
interested in the influence of a finite damping strength. In the following, we will 
concentrate on the difference $S-S_0$ which is independent of $L$.

For high temperatures, one finds from (\ref{eq:s_fp})
\begin{equation}
\label{eq:s_fp_highT}
\frac{S-S_0}{k_B} = \frac{1}{2}\left[\log\left(\frac{k_BT}{\hbar\gamma}\right)+1\right]\,.
\end{equation}
In view of the entropy constant (\ref{eq:s0}) it follows that the high-temperature
behavior of the entropy $S$ does not depend on the damping strength $\gamma$. This 
reflects the fact that equilibrium properties in classical thermodynamics do not 
depend on the strength of the coupling between system and heat bath. At the same time,
the high-temperature expression for the entropy represents the entropy of an undamped
free particle for arbitrary temperatures $T>0$. By means of the relation 
\begin{equation}
C=T\frac{\partial S}{\partial T}
\end{equation}
between specific heat $C$ and entropy $S$, the logarithmic dependence of the entropy
on the inverse temperature leads to a constant specific heat $C=k_B/2$ for any 
nonvanishing temperature in the absence of damping. 

In the presence of damping, quantum effects arise in the low-temperature regime where 
(\ref{eq:s_fp}) yields
\begin{equation}
\label{eq:s_fp_lowT}
\frac{S-S_0}{k_B} = \frac{\pi}{3}\frac{k_BT}{\hbar\gamma}\,.
\end{equation}
This expression diverges for vanishing damping constant, clearly indicating
that the limit $\gamma\to0$ is nontrivial at low temperatures. In contrast to
the constant specific heat found for the undamped particle, the specific heat
now depends linearly on temperature.

\begin{figure}
\includegraphics[width=\columnwidth]{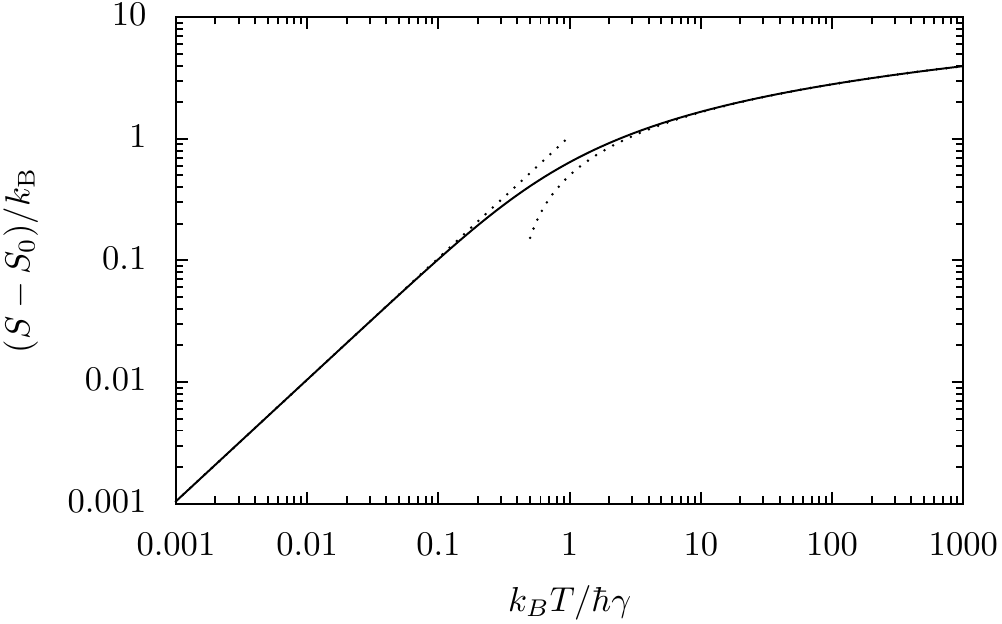}
\caption{Temperature dependence of the difference between the entropy (\ref{eq:s_fp}) 
and its zero-temperature value (\ref{eq:s0}) for a free Brownian particle 
subject to ohmic dissipation of strength $\gamma$. The dotted lines correspond to the
high- and low-temperature expressions (\ref{eq:s_fp_highT}) and (\ref{eq:s_fp_lowT}),
respectively.}
\label{fig:entropy_fp} 
\end{figure}

The temperature dependence of the difference between the entropy
(\ref{eq:s_fp}) and its zero-temperature value (\ref{eq:s0}) is depicted in
Fig.~\ref{fig:entropy_fp} as solid line together with the high- and
low-temperature expressions (\ref{eq:s_fp_highT}) and (\ref{eq:s_fp_lowT}),
respectively, as dotted lines. 

From (\ref{eq:s_fp}) it is clear, that the difference $S-S_0$ depends on the
temperature only through the dimensionless quantity $\hbar\beta\gamma$ which,
in view of our discussion in the introduction, should be expected. As $\gamma$
decreases, the high-temperature region effectively becomes increasingly larger.
However, the divergence of (\ref{eq:s_fp_highT}) in the zero-temperature limit
is avoided by a crossover to the low-temperature behavior (\ref{eq:s_fp_lowT})
below temperatures of the order of $\hbar\gamma/k_B$. In this low-temperature
regime, the dependence on $\hbar\beta\gamma$ immediately implies a nontrivial
limit of vanishing coupling to the heat bath as discussed in connection with
(\ref{eq:s_fp_lowT}).

As we will see in more detail in the discussion of the Casimir entropy, a
difference between the Casimir problem and the free Brownian particle consists
in the value taken by the entropy at zero temperature. While the Casimir
entropy goes to zero in that limit, we found a nonvanishing entropy constant
$S_0$ for the free Brownian particle. $S_0$ will be positive provided that the
broadening of the levels due to the coupling to the heat bath is larger than
the ground-state energy. The Casimir entropy, on the other hand, can exhibit
negative values of the entropy under certain circumstances. This raises the
question whether in the case of the free Brownian particle the entropy can drop
below $S_0$. While this does not necessarily imply a negative entropy, the
entropies in the two situations would nevertheless behave nonmontonically. As a
direct consequence, measurable thermodynamic quantities like the specific heat
could in both cases take on negative values. This was indeed found for the free
Brownian particle \cite{hangg08}.

\begin{figure}
\includegraphics[width=\columnwidth]{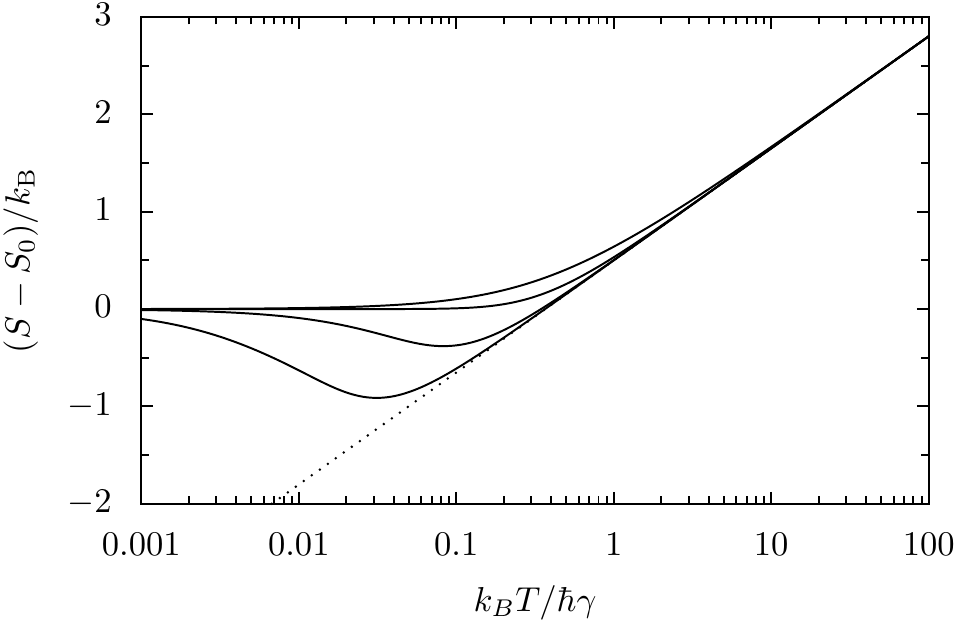}
\caption{Temperature dependence of the difference between the entropy (\ref{eq:s_fp}) 
and its zero-temperature value (\ref{eq:s0}) for a free Brownian particle 
subject to ohmic dissipation with a high-frequency cutoff $\omega_c/\gamma=\infty,1,0.1$, and 
$0.01$ from the upper to the lower curve. The dotted line represents the high-temperature
expression (\ref{eq:s_fp_highT}).}
\label{fig:entropy_fp_drude} 
\end{figure}

For strictly ohmic damping, $\hat\gamma(\xi) = \gamma$, the entropy $S$ will
always be larger than $S_0$ and we therefore have to consider the case of
finite cutoff frequency $\omega_c$.  If $\hat\gamma'(0)<-1$, which for the
damping kernel (\ref{eq:drudekernel}) corresponds to $\omega_c<\gamma$, there
exists a temperature range where $S-S_0<0$ and where the specific heat becomes
negative \cite{hangg08}. The nonmonotonic behavior of the entropy is shown in
Fig.~\ref{fig:entropy_fp_drude} for decreasing values of the cutoff frequency
$\omega_c/\gamma =\infty, 1, 0.1,$ and $0.01$ from the upper to the lower
curve. In addition to the appearance of a nonmonotonic behavior, one observes
that with decreasing cutoff frequency $\omega_c$ and therefore decreasing
coupling to the bath oscillators, the entropy follows the high-temperature
behavior (\ref{eq:s_fp_highT}) represented by the dotted line down to lower
temperatures. This underlines once more the nontrivial limit of vanishing
coupling to the heat bath.

The nonmonotonicity of the entropy however does not put the thermodynamic
stability of the system in question. As pointed out before, an entropy based on
the reduced partition function (\ref{eq:partfunc}) is the difference of two
positive entropies referring to the heat bath and the heat bath in the presence
of the free particle \cite{ingol09}. 

\section{Plasma and Drude models}
\label{sec:drudemodel}

In the context of this paper, we are not interested in a realistic description
of experiments \cite{decca07,munda07,chan08,vanzw08,munda09,jourd09,masud09,deman09}.
exploring the Casimir effect but rather in a theoretical analysis of the
effects of a finite conductivity of the mirrors, in particular at low
temperatures. We therefore assume an idealized geometrical setup where two
infinitely large metallic plane mirrors are positioned in parallel at a
distance $L$ of each other.  By an appropriate choice of the permittivity, we
will account for basic features of the optical response of real mirrors, namely
the low reflectivity at high frequencies and the finite conductivity at low
frequencies.

The high-frequency optical response of a metal is well described within the plasma 
model in terms of the relative permittivity 
\begin{equation}
\label{eq:epsilon_plasma}
\varepsilon(\omega) = 1-\frac{\omega_P^2}{\omega^2}\,.
\end{equation}
Here, $\omega_P$ denotes the plasma frequency which is related to the plasma
wavelength by
\begin{equation}
\label{eq:lambda_p}
\lambda_P = \frac{2\pi c}{\omega_P}
\end{equation}
where $c$ is the speed of light. For gold, the plasma wavelength is $0.136\,\mu$m
and the finite plasma frequency becomes appreciable for mirror distances 
$L\lesssim1\,\mu$m \cite{genet00}. In the following, we assume for all numerical 
calculations a fixed value of $\lambda_P/L=0.136$, corresponding to a mirror 
distance of one micrometer which is a typical order of magnitude in experiments.

The relative permittivity $\varepsilon(\omega)$ is related to the conductivity 
$\sigma(\omega)$ of the mirrors by
\begin{equation}
\label{eq:epsilon_sigma}
\varepsilon(\omega) = 1+i\frac{\sigma(\omega)}{\omega}\,.
\end{equation}
Here, the conductivity $\sigma$ is expressed as a frequency from which the
conductivity in SI units is obtained as $\varepsilon_0\sigma$ with the
permittivity of vacuum $\varepsilon_0$. According to (\ref{eq:epsilon_sigma}),
applying the relative permittivity (\ref{eq:epsilon_plasma}) for all
frequencies down to $\omega=0$ would imply the assumption of an infinite dc
conductivity which is certainly unacceptable for normal metals. 

The simplest model ensuring a finite dc conductivity is the Drude model with
the relative permittivity
\begin{equation}
\label{eq:epsilon_drude}
\varepsilon(\omega) = 1-\frac{\omega_P^2}{\omega(\omega+i\gamma)}\,.
\end{equation}
It is obtained from (\ref{eq:epsilon_plasma}) by introducing a relaxation 
frequency $\gamma$ and leads to the dc conductivity 
\begin{equation}
\label{eq:sigma0}
\sigma_0 = \frac{\omega_P^2}{\gamma}\,.
\end{equation}
We will assume $\gamma$ to be independent of temperature thus implying a 
nonvanishing conductivity at zero temperature due to crystal defects or 
impurities \cite{tto0}. 

Taking the dc conductivity of gold, $\varepsilon_0\sigma_\text{Au} =
4.52\cdot10^7\,(\Omega\text{m})^{-1}$, one obtains for the relaxation frequency
$\gamma\approx 2.7\cdot 10^{-3}\omega_P$.  Assuming mirror distances of one
micrometer or smaller, we find $\gamma L/c\lesssim 0.125$.  The relaxation
frequency therefore is typically small compared to $\omega_p$ and to $L/c$. On
the other hand, for the conductivity assumed here, the ratio $\hbar\gamma/k_B
T$ is of the order of one at room temperature.

Although the Drude model cannot be expected to give an accurate description of
all details of the permittivity of a real metal \cite{lambr00,sveto08}, it is the simplest 
model accounting for a finite dc conductivity and therefore deserves to be analyzed 
in detail. As apart from the thermal frequency $k_BT/\hbar$ the relaxation frequency 
$\gamma$ is the smallest frequency scale in the problem, it is interesting to 
investigate the limit $\gamma\to0$. 

The reason why this limit can be nontrivial becomes apparent when the
reflection coefficients are considered. Before doing so, we introduce the
notation for frequencies and wavevectors which will be used in the sequel. The
wavevector of an electromagnetic mode is
$\mathbf{k}=(\mathbf{k}_\parallel,k_\bot)$ where $k_\bot$ denotes the component
orthogonal to the mirrors while $\mathbf{k}_\parallel$ is a two-dimensional
vector parallel to the mirrors. Although it is possible and sometimes useful
to keep frequencies real \cite{sveto07,ellin08}, here we will employ imaginary
frequencies $\xi=-i\omega$ and wave-vector components $\kappa=-ik_\bot$ so that
the dispersion relation becomes $\xi^2=c^2(\kappa^2-\mathbf{k}_\parallel^2)$.

With this notation the reflection coefficient for the transverse electric (TE) 
mode is obtained as
\begin{equation}
\label{eq:r_te}
r_\text{TE}(i\xi,i\kappa)=\frac{\sqrt{c^2\kappa^2+[\varepsilon(i\xi)-1]\xi^2}
-c\kappa}{\sqrt{c^2\kappa^2+[\varepsilon(i\xi)-1]\xi^2}+c\kappa}
\end{equation}
while for the transverse magnetic (TM) mode one finds
\begin{equation}
\label{eq:r_tm}
r_\text{TM}(i\xi,i\kappa)=\frac{\sqrt{c^2\kappa^2+[\varepsilon(i\xi)-1]\xi^2}
-\varepsilon(i\xi)c\kappa}{\sqrt{c^2\kappa^2+[\varepsilon(i\xi)-1]\xi^2}+
\varepsilon(i\xi)c\kappa}\,.
\end{equation}
An important difference between the plasma model and the Drude model with
direct implications for the reflection coefficients is the fact that the limit
of $\xi^2[\varepsilon(i\xi)-1]$ for $\xi\to0$ tends to $\omega_P^2$ for the
plasma model while it vanishes for the Drude model. This does not affect the
zero frequency reflection of the TM mode but for the TE mode one finds a
vanishing reflection coefficient $r_\text{TE}(0,i\kappa)$ for the Drude model
in contrast to a nonvanishing value for the plasma model \cite{bostr00}.

Clearly, such a behavior is not compatible with a continuous transition from
the Drude to the plasma model for $\gamma\to0$. The zero-frequency behavior of
the reflection coefficient has a dramatic consequence for the Casimir force at
high temperatures. While within the plasma model the TE modes contribute to the
Casimir force, this is not the case within the Drude model.  As a consequence,
in the latter case the Casimir force arises only from the TM modes and is thus
reduced by a factor of two with respect to the plasma model \cite{bostr00}.
Recent years have seen an extensive debate about the existence of this
reduction factor and its compatibility with thermodynamics (for a review see
\textit{e.g.}\ Refs.~\onlinecite{klimc06} and \onlinecite{milto08}).
Interestingly, a rather independent approach based on a microscopic study of
the force between two slabs containing an electron plasma in the classical and
semiclassical regime also led to a reduction of the Casimir force by a factor
of two \cite{janco05,buenz05a} as does the consideration of the classical
Bohr-van Leeuwen theorem \cite{bimon09}. 

In the following, we will analyze in detail the transition from the Drude
to the plasma model. Of particular interest will be the case of low
temperatures. This limit is of relevance for the leading thermal
corrections to the Casimir force and also in the context of the ongoing
debate about a possible violation of the third law of thermodynamics
\cite{klimc06,milto08}. In the course of the discussion it will become clear
that the transition from the Drude to the plasma model for the TE mode 
bears close analogy to the transition from a free Brownian particle to a 
free particle \cite{hangg06,hangg08,ingol09}.

\section{Thermal corrections to the Casimir force}
\label{sec:thermalcorr}

For the case of only partially reflecting mirrors, the radiation pressure of
the electromagnetic field modes on the mirrors can be obtained from a scattering 
formalism \cite{jaeke91}. Performing the integration over the modes in imaginary
frequency $\xi$ and imaginary orthogonal component $\kappa$ of the wavevector, 
one finds for the Casimir force between two parallel mirrors \cite{genet03}
\begin{equation}
\label{eq:f_resummed}
F=\frac{\hbar A}{\pi^2}\sum_{n=0}^\infty\strut'\int_0^\infty d\xi
\cos(n\hbar\beta\xi)\mathbb{F}(\xi)\,.
\end{equation}
Here, $A$ is the surface of the mirror, the prime at the sum sign indicates
that the $n=0$ term contributes only with a factor one-half, and
\begin{equation}
\label{eq:f_xi}
\mathbb{F}(\xi) = \int_{\xi/c}^\infty d\kappa\,\kappa^2 f(i\xi,i\kappa) \,.
\end{equation}
In the integrand\,,
\begin{equation}
\label{eq:closedloopfunction}
f(i\xi,i\kappa) = \sum_{p=\mathrm{TE, TM}}f_p(i\xi,i\kappa)
\end{equation}
is the closed-loop function where one has to sum over the two polarizations $p$
with
\begin{equation}
\label{eq:clp_per_mode}
f_p(i\xi,i\kappa) = \frac{r_p^2(i\xi,i\kappa)}{\exp(2\kappa L)-r_p^2(i\xi,i\kappa)}\,.
\end{equation}
In this expression, we have assumed for simplicity that the two mirrors have the 
same reflection properties.

The $n=0$ term in (\ref{eq:f_resummed}) corresponds to the zero-temperature
Casimir force arising from the vacuum fluctuations of the electromagnetic field
between the mirrors. All terms with $n>0$ describe thermal corrections due
to real photons present at finite temperatures. Equation (\ref{eq:f_resummed})
is therefore particularly well suited to separate off the zero-temperature force
and to determine the low-temperature behavior.

If the conditions of validity of the Poisson formula are met \cite{milto04,reyna04},
we recover from (\ref{eq:f_resummed}) the Lifshitz formula of the Casimir force
\cite{lifsh56}
\begin{equation}
\label{eq:f_lifshitz}
F = \frac{A}{\pi\beta}\sum_{n=0}^\infty\strut'\mathbb{F}(\xi_n)\,,
\end{equation}
where $\mathbb{F}(\xi)$ has been defined in (\ref{eq:f_xi}) and $\xi_n$ are
the Matsubara frequencies (\ref{eq:matsubarafreq}).  Here, the $n=0$ term
allows to immediately read off the dominant contribution to the force at high
temperatures. For the TE mode within the Drude model, the reflection
coefficient and therefore the closed-loop function vanish at $\xi=0$. According
to (\ref{eq:f_lifshitz}), the TE modes thus do not contribute to the
high-temperature behavior, resulting in the reduction of the force by a factor
of two mentioned above \cite{bostr00}. The expression (\ref{eq:f_lifshitz}) is
not restricted to high temperatures though and a corresponding expression has
recently been employed to analyze the low-temperature behavior of the free
energy \cite{hoye07,brevi08,ellin08a}.

For the discussion of the thermal corrections it is convenient to introduce a 
dimensionless factor by dividing the Casimir force $F$ through the Casimir 
force for ideal mirrors at zero temperature
\begin{equation}
\label{eq:f_ideal}
F_\text{Cas} = \frac{\hbar cA\pi^2}{240L^4}\,.
\end{equation}
This expression accounts for both the TE and the TM mode and will be used
even in cases where only one of the modes is considered. We thus define the
factor
\begin{equation}
\label{eq:eta_f}
\eta_F = \frac{F}{F_\text{Cas}} = \eta_F^0+\eta_F^T
\end{equation}
where $\eta_F^0$ is the zero-temperature contribution arising from the $n=0$ term 
in (\ref{eq:f_resummed}) and 
\begin{equation}
\label{eq:eta_f_t}
\eta_F^T = \frac{240L^4}{c\pi^4}\sum_{n=1}^\infty\int_0^\infty d\xi
\cos(n\hbar\beta\xi)\mathbb{F}(\xi)
\end{equation}
describes the thermal corrections. 

\begin{figure}
\includegraphics[width=\columnwidth]{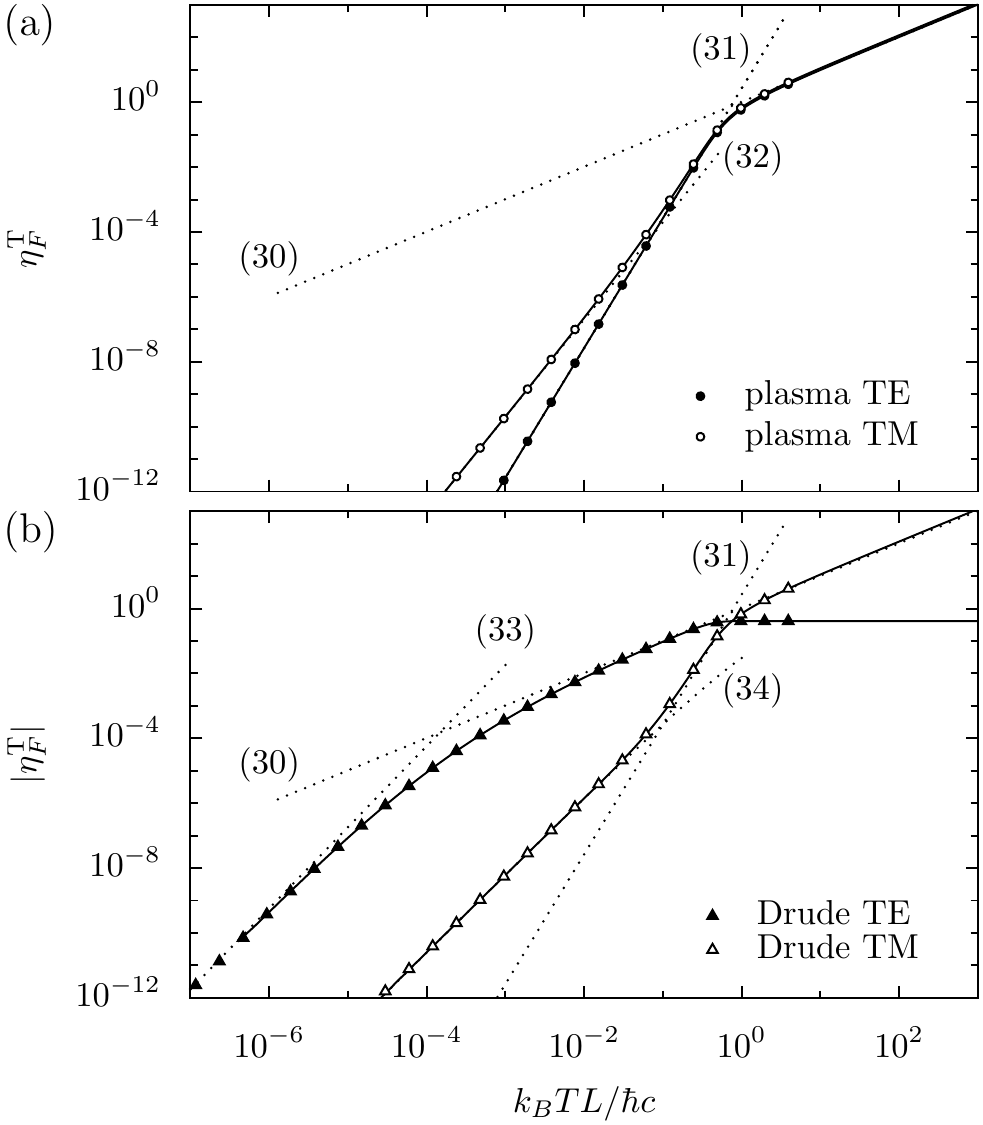}
\caption{The thermal contribution to the factor $\eta_F$ defined in (\ref{eq:eta_f})
for (a) the plasma model and (b) the Drude model is shown in terms of the 
dimensionless factor (\ref{eq:eta_f_t}) as a function of the temperature 
for $\lambda_P/L=0.136$ and $\sigma_0=\sigma_\text{Au}$. The solid lines were obtained 
from the Lifshitz formula (\ref{eq:f_lifshitz}) while the filled and open symbols
were obtained from (\ref{eq:f_resummed}) for the TE and TM mode, respectively. 
The thermal contribution of the TE mode within the Drude model is negative so
that the absolute value of the data is shown. The dotted lines correspond
to various low- and high-temperature approximations identified by the 
respective equation numbers. In (\ref{eq:lowTdrudeTM}), $C=-2.58$ was used.}
\label{fig:force_thermal} 
\end{figure}

Before entering into a detailed analysis of the TE mode in Sec.~\ref{sec:temode}, 
we discuss the temperature dependence of the Casimir force for the TE and TM modes 
in the plasma model and the Drude model. Figure~\ref{fig:force_thermal} displays 
the thermal contribution $\eta_F^T$ defined in (\ref{eq:eta_f_t}) as a function of 
the dimensionless temperature $k_BTL/\hbar c$ for $\lambda_P/L=0.136$ and
the dc conductivity of gold $\sigma_\text{Au}$. The solid lines represent 
results based on the Lifshitz formula (\ref{eq:f_lifshitz}) while the filled and open 
symbols represent data obtained by evaluating (\ref{eq:f_resummed}) for the TE and 
the TM mode, respectively. As Fig.~\ref{fig:force_thermal} indicates, the two approaches
(\ref{eq:f_resummed}) and (\ref{eq:f_lifshitz}) lead indeed to the same results,
confirming that the Poisson resummation can also be employed for the Drude model.
For the numerical evaluation of the force, the expression (\ref{eq:f_resummed}) works
well at low temperatures while (\ref{eq:f_lifshitz}) is advantageous at higher
temperatures.

Results for the plasma model are shown in Fig.~\ref{fig:force_thermal}a. The 
dotted lines indicate the high-temperature approximation
\begin{equation}
\label{eq:highTapprox}
\eta_F^T = \frac{120}{\pi^3}L^3\mathbb{F}(0)\frac{k_BTL}{\hbar c}
\end{equation}
as well as the low-temperature approximations for the TE mode
\begin{equation}
\label{eq:lowTplasmaTE}
\eta_F^T = \frac{8}{3}\left(\frac{k_BTL}{\hbar c}\right)^4
\end{equation}
and for the TM mode \cite{borda00,bezer02}
\begin{equation}
\label{eq:lowTplasmaTM}
\eta_F^T = \frac{240}{\pi^3}\zeta(3)\frac{c}{\omega_PL}\left(\frac{k_BTL}{\hbar c}\right)^3\,,
\end{equation}
where the value of the Riemann zeta function is $\zeta(3)=1.202\ldots$ The expression
(\ref{eq:lowTplasmaTE}) is independent of the plasma frequency and thus agrees with
the leading thermal correction in the case of ideal mirrors \cite{sauer62,mehra67,schwi78}.

The temperature dependence of the Casimir force within the Drude model is depicted
in Fig.~\ref{fig:force_thermal}b. The low-temperature approximations again shown
as dotted lines differ from those of the plasma model. For the TE mode one finds
\begin{equation}
\label{eq:lowTdrudeTE}
\eta_F^T = -\frac{15}{\pi^4}\left(\frac{\pi}{2}\right)^{1/2}\zeta\left(\frac{5}{2}\right)
\left(\frac{L\sigma_0}{c}\right)^{3/2}\left(\frac{k_BTL}{\hbar c}\right)^{5/2}\,,
\end{equation}
where $\zeta(5/2)=1.341\ldots$ The dependence on the material properties of the
mirrors, \textit{i.e.}\ in our case the plasma frequency and the relaxation
frequency, enter here only through the dc conductivity (\ref{eq:sigma0}). The
fact that the prefactor increases with increasing conductivity indicates that
the limit from the Drude to the plasma model is indeed nontrivial for the TE
mode. This behavior is analogous to the low-temperature behavior
(\ref{eq:s_fp_lowT}) of the entropy of a free damped particle where the
prefactor increases with decreasing damping constant.

The thermal correction (\ref{eq:lowTdrudeTE}) is negative and thereby reduces
the zero-temperature Casimir force, thus already hinting at the fact that the
TE modes will no longer contribute to the force at high temperatures.
Accordingly, the thermal contribution $\eta_F^T$ for this mode saturates at
$-\eta_F^0$, \textit{i.e.}\ the negative of the zero-temperature contribution,
so that the total force due to the TE modes vanishes in the high-temperature
limit. 

The expression (\ref{eq:lowTdrudeTE}) confirms the expression appearing as
next-to-leading order in the low-temperature expansion of the free energy
conjectured by H{\o}ye \textit{et al.} \cite{hoye07} on the basis of an
analysis of the Lifshitz formula \cite{footnote2} and later derived by Borel
summation \cite{ellin08a}. We will turn to the low-temperature behavior of the
free energy in Sec.~\ref{sec:entropy}.

The low-temperature approximation for the TM mode
\begin{equation}
\begin{aligned}
\label{eq:lowTdrudeTM}
\eta_F^T &= \frac{90}{\pi^4}(2\pi)^{1/2}\zeta\left(\frac{5}{2}\right)
\left(\frac{c}{L\sigma_0}\right)^{1/2}\left(\frac{k_BTL}{\hbar c}\right)^{5/2}\\
&\quad\times\left[\log\left(\hbar\omega_P/k_BT\right)+C\right]
\end{aligned}
\end{equation}
is special because instead of a simple power law a logarithmic factor appears.
$C$ is a constant for which (\ref{eq:constant_tm}) represents one contribution.
As in (\ref{eq:lowTdrudeTE}), the material properties of the mirror appear in the 
prefactor only through the dc conductivity (\ref{eq:sigma0}). In contrast to the
TE mode, however, the prefactor decreases here with increasing conductivity so that
the limit of infinite conductivity does not present difficulties. A derivation of 
(\ref{eq:lowTdrudeTE}) and (\ref{eq:lowTdrudeTM}) can be found in the appendix. 

After multiplication of the low-temperature approximations (\ref{eq:lowTplasmaTE}), 
(\ref{eq:lowTplasmaTM}), (\ref{eq:lowTdrudeTE}), and (\ref{eq:lowTdrudeTM})
with the Casimir force for ideal mirrors at zero temperature (\ref{eq:f_ideal})
one notices that the leading thermal corrections to the force do not depend
on the distance $L$ between the mirrors for the TE mode while they decrease 
as $1/L^2$ for the TM mode. The fact that the leading thermal correction for
the TE mode is independent of $L$ does not mean, however, that at a fixed
low temperature this contribution will survive for arbitrarily large separations
of the two mirrors because an increase in $L$ will drive the system into the
regime where the high-temperature approximation (\ref{eq:highTapprox}) applies.
Then the Casimir force decreases as the mirrors are moved apart.

One feature visible in Fig.~\ref{fig:force_thermal} for the parameters chosen
here is still worth being noted. Both for the plasma model and the Drude model,
the thermal contribution of the TM modes in an intermediate temperature regime
increases with temperature according to the low-temperature behavior
(\ref{eq:lowTplasmaTE}) of the TE modes for ideal mirrors.  This happens
because the thermal correction for the TM modes increases more slowly with
temperature than for the TE mode in the presence of ideal mirrors. Before the
high-temperature asymptote is reached, the thermal correction crosses over to
the low-temperature behavior (\ref{eq:lowTplasmaTE}) of the TE mode in the
ideal case or the plasma model.  In this regime, the thermal corrections are
dominated by the exponential factor in the closed-loop function
(\ref{eq:clp_per_mode}) and the reflection constant can be set to one. 

\begin{figure}
\includegraphics[width=\columnwidth]{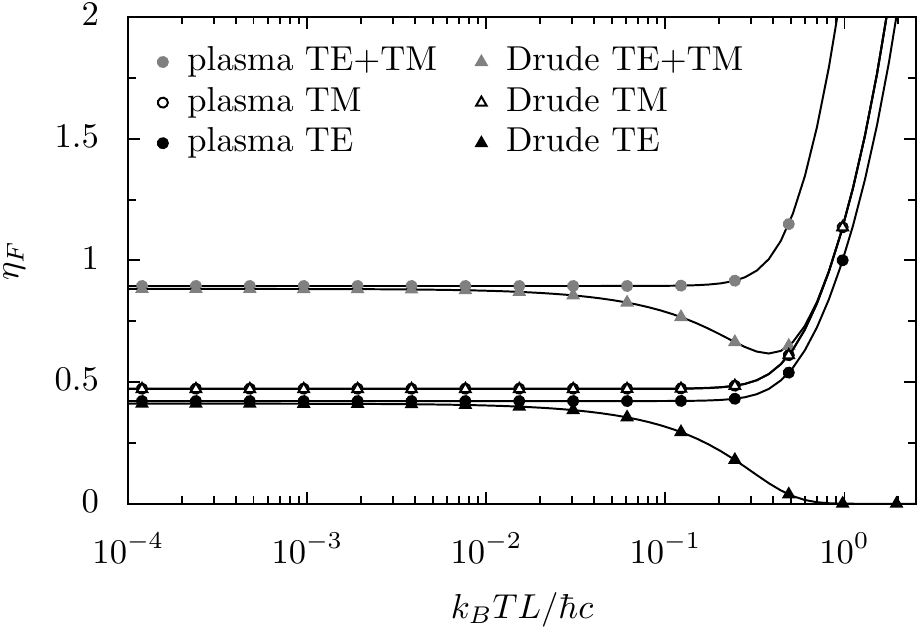}
\caption{The total Casimir force is shown in terms of the dimensionless factor
defined in (\ref{eq:eta_f}) as a function of the temperature for
$\lambda_P/L=0.136$ and $\sigma_0=\sigma_\text{Au}$. The lines were obtained from the
Lifshitz formula (\ref{eq:f_lifshitz}) while the symbols represent data obtained
from (\ref{eq:f_resummed}) for the plasma model (circles) and the Drude model
(triangles). Filled and open symbols correspond to TE and TM modes,
respectively, while grey symbols represent the sum of both modes. For the TM
mode, because of the small relaxation frequency, the curves and symbols for the 
two models lie almost exactly on top of each other.}
\label{fig:force_total} 
\end{figure}

In Fig.~\ref{fig:force_total} we show the total force expressed through the
factor $\eta_F$ as a function of the temperature for $\lambda_P/L=0.136$ and
the dc conductivity of gold. As in Fig.~\ref{fig:force_thermal} the solid lines
result from an evaluation of the Lifshitz formula (\ref{eq:f_lifshitz}) while
the symbols represent data obtained by means of (\ref{eq:f_resummed}). Filled
and open symbols correspond to TE and TM modes, respectively, while the grey
symbols represent the sum of both. Circles and triangles correspond to the
plasma and the Drude model, respectively. As noted before, the agreement
between the two approaches (\ref{eq:f_resummed}) and (\ref{eq:f_lifshitz})
confirms the applicability of the Poisson resummation. 

For the relatively small value of the relaxation frequency $\gamma$ chosen
here, it is almost impossible for the TM modes to distinguish in
Fig.~\ref{fig:force_total} between the data corresponding to the Drude and
plasma models. For these modes, the limit $\gamma\to0$ continuously leads from the
Drude to the plasma model. In contrast, the contribution of the TE modes
behaves very differently for the Drude model and the plasma model. This
difference survives the limit $\gamma\to0$.

So far, an experimental distinction between the plasma and the Drude model is
only based on an effectively zero-temperature measurement of the Casimir force
\cite{decca07}. This quantity integrates over the closed-loop function and one
therefore has to rely on quantitative comparisons. Thermal contributions to the
Casimir force at low temperatures, on the other hand, are particularly
sensitive to the low-frequency behavior of the closed-loop function and allow
for a decision in favor of one or the other model already on a qualitative
level. One of the indicators would evidently be the sign of the thermal
correction. In addition, because of the different exponent in the
low-temperature expressions (\ref{eq:lowTplasmaTE}) and (\ref{eq:lowTdrudeTE})
of the TE mode, thermal corrections for the Drude model should be visible
already at lower temperatures or smaller mirror separation than expected for 
the plasma model.

\section{The transverse electric mode}
\label{sec:temode}

We will now focus on the TE mode. To avoid cumbersome notation, it will 
be understood that all quantities refer to the TE mode even if this 
is not made explicit. In order to analyze the difference in behavior for
the plasma model on the one hand and the Drude model on the other hand,
it is appropriate to consider the difference in the thermal contributions 
to the factor $\eta_F$ defined in (\ref{eq:eta_f})
\begin{equation}
\label{eq:d_eta_f_t}
\Delta\eta_F^T = \eta_F^T(\text{Drude})-\eta_F^T(\text{plasma})\,.
\end{equation} 

To understand the transition from the Drude to the plasma model, it is crucial
to realize that the closed-loop function $f(i\xi,i\kappa)$ for the Drude model
depends only on the ratio $\xi/\gamma$. According to (\ref{eq:r_te}) and
(\ref{eq:clp_per_mode}), the frequency enters only through the combination
$\xi^2[1-\varepsilon(i\xi)]$, so that our assertion immediately follows from
the relative permittivity (\ref{eq:epsilon_drude}) of the Drude model.

\begin{figure}
\includegraphics[width=\columnwidth]{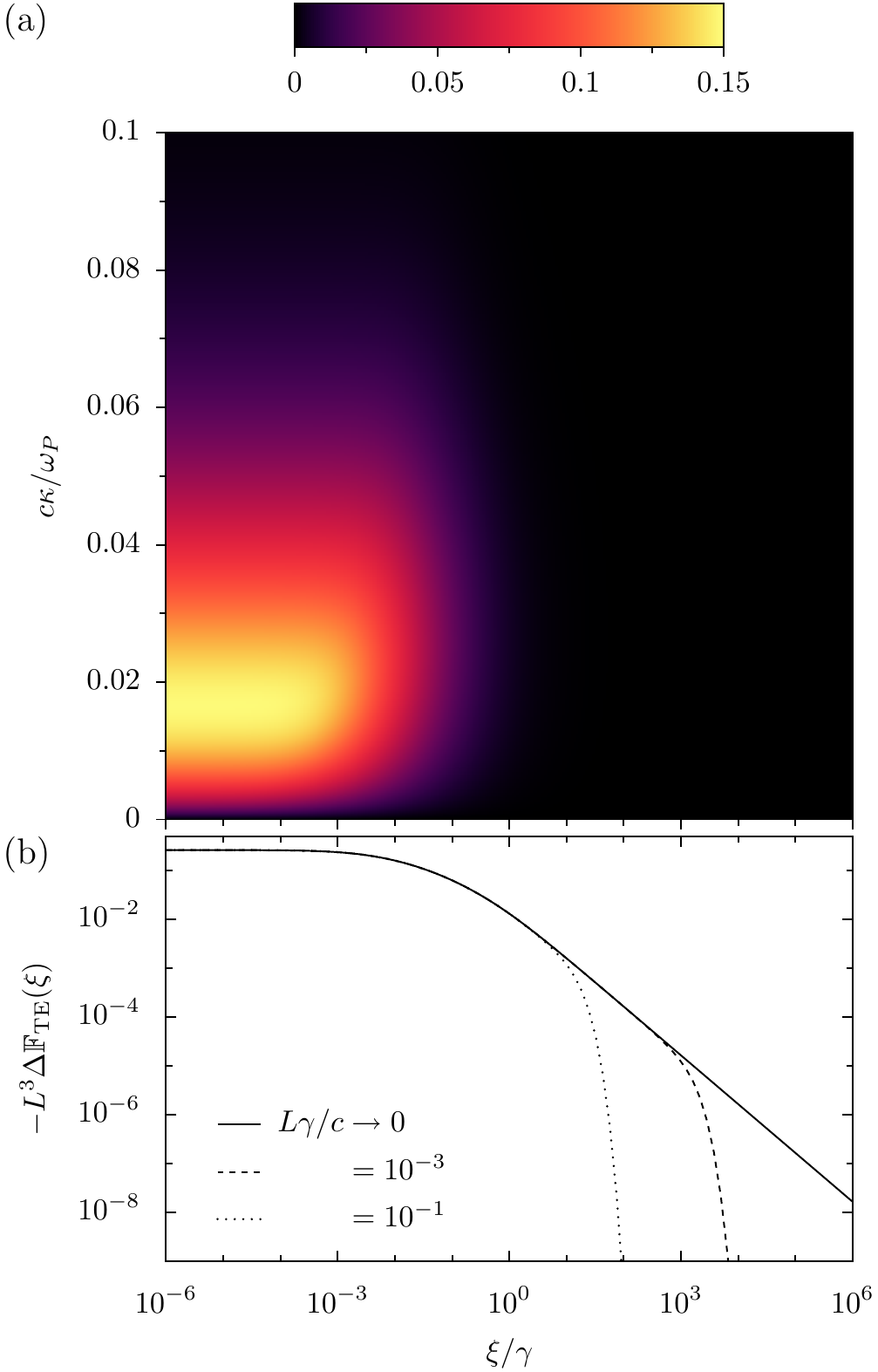}
\caption{(a) The difference $-(\kappa L)^2\Delta f_\text{TE}$ between the 
integrands appearing in (\ref{eq:f_xi}) for the Drude and the plasma model 
is shown as a function of $\xi/\gamma$ and $c\kappa/\omega_P$ for 
$\lambda_P/L=0.136$. (b) After integration over $\kappa$, one finds 
$-L^3\Delta\mathbb{F}_\text{TE}(\xi)$ which tends to the solid line in the 
limit $\gamma\to0$. The dashed and dotted lines refer to $L\gamma/c=10^{-3}$ 
and $10^{-1}$, respectively.}
\label{fig:clp_te} 
\end{figure}

This behavior is illustrated in Fig.~\ref{fig:clp_te} where in the upper panel
$-(\kappa L)^2\Delta f_\text{TE}$ is plotted as a function of
$c\kappa/\omega_P$ and $\xi/\gamma$. In analogy to (\ref{eq:d_eta_f_t}), $\Delta
f_\text{TE}$ refers to the difference in the closed-loop functions of the TE
mode within the Drude and the plasma model. The plotted quantity then
describes the change of the integrand in (\ref{eq:f_xi}) when going from the 
plasma model to the Drude model.

Figure~\ref{fig:clp_te}b depicts $-L^3\Delta\mathbb{F}_\text{TE}(\xi)$ which
according to (\ref{eq:f_xi}) is obtained from the quantity shown in
Fig.~\ref{fig:clp_te}a by integration over $\kappa$. Due to the lower integration 
limit, $\mathbb{F}_\text{TE}$ for the Drude model and thus 
$\Delta\mathbb{F}_\text{TE}$ depend on $\gamma$. The dotted and dashed 
lines correspond to $L\gamma/c=10^{-1}$ and $10^{-3}$, respectively, while the
solid line represents the limit $\gamma\to0$. In agreement with the reasoning above,
the low-frequency region, where this function has its largest weight, turns out
to be practically independent of $\gamma$ as long as the relaxation frequency
is not too large. Only at higher frequencies, finite relaxation frequencies
lead to a rather sharp cutoff. With decreasing relaxation frequency, also the
high-frequency part approaches more and more the decay with $1/\xi$ found in
the limit $\gamma\to0$.

In view of this observation, it is appropriate to rescale the frequency in
(\ref{eq:eta_f_t}) by $\gamma$ which implies that the temperature only
appears in the combination $\hbar\beta\gamma$. The change (\ref{eq:d_eta_f_t})
in the thermal contribution to the factor $\eta_F$ then turns into
\begin{equation}
\label{eq:delta_eta_f_t}
\begin{split}
\Delta\eta_F^T &= \frac{240}{\pi^4}\left(\frac{L\omega_P}{c}\right)^3
\frac{L\gamma}{c}\sum_{n=1}^\infty\int_0^\infty dx
\cos(n\hbar\beta\gamma x)\\
&\quad\times\int_{(\gamma/\omega_P)x}^\infty du\,u^2
\left[f^D\left(i\gamma
x,i\frac{\omega_P}{c}u\right)-f^P\left(i\frac{\omega_P}{c}u\right)\right]
\end{split}
\end{equation}
where $u=c\kappa/\omega_P$ and $f^D$ and $f^P$ are the closed-loop functions
(\ref{eq:clp_per_mode}) with the reflection coefficients of the Drude
and plasma model, respectively.

In the limit of small $\gamma$, the lower limit of the inner integral in 
(\ref{eq:delta_eta_f_t}) effectively becomes zero. Then, the difference
of the thermal corrections can be expressed as
\begin{equation}
\label{eq:delta_eta_ft_scaling}
\Delta\eta_F^T = \gamma g_F(\hbar\beta\gamma)\,.
\end{equation}
In the special cases of Eqs.~(\ref{eq:highTapprox}) and (\ref{eq:lowTdrudeTE})
we have $g_F(x)\sim 1/x$ and $g_F(x)\sim 1/x^{5/2}$, respectively. The prefactor
$\gamma$ in (\ref{eq:delta_eta_ft_scaling}) ensures that the sum appearing in 
(\ref{eq:delta_eta_f_t}) has a well-defined high-temperature limit.

As the temperature appears in (\ref{eq:delta_eta_ft_scaling}) only in the 
combination $\hbar\beta\gamma$, it follows that the limits of zero temperature 
and zero relaxation frequency do not commute. For any finite temperature, the 
limit $\gamma\to0$ implies $\hbar\beta\gamma\to0$ so that (\ref{eq:d_eta_f_t}) 
takes its classical value
\begin{equation}
\label{eq:delta_eta_f_t0}
\Delta\eta_F^T(\gamma=0) = -\frac{120}{\pi^3}\left(\frac{L\omega_P}{c}\right)^3
\frac{L}{\hbar\beta c}\int_0^\infty du\,u^2
f_P\left(i\frac{\omega_P}{c}u\right)\,.
\end{equation}
This result is independent of the closed-loop function of the Drude model
because the reflection coefficient within this model vanishes at zero
frequency. 

\begin{figure}
\includegraphics[width=\columnwidth]{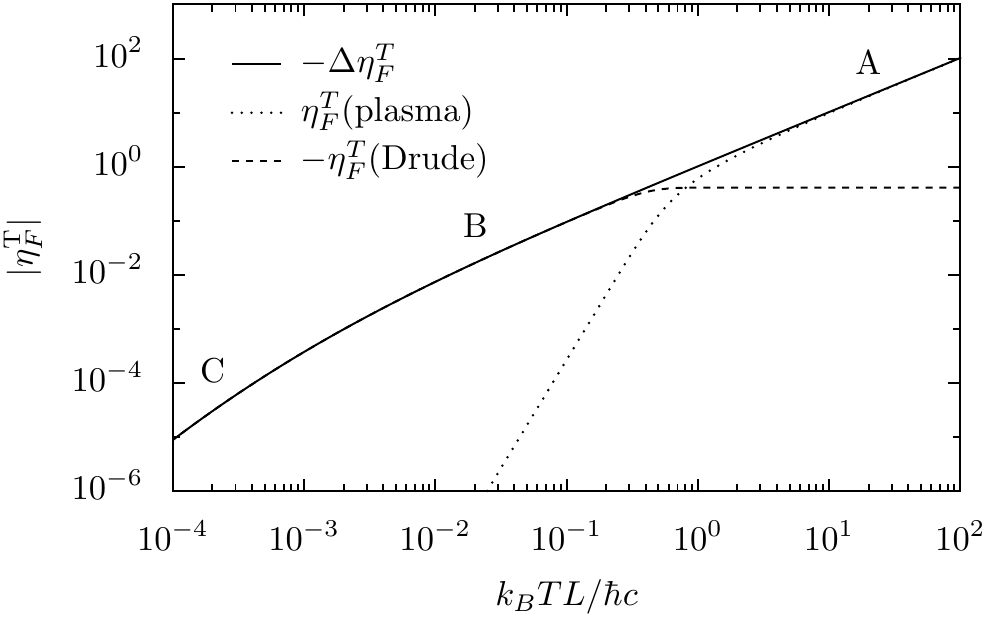}
\caption{The absolute value of the difference of the thermal corrections 
(\ref{eq:delta_eta_f_t}) to the factor $\eta_F$ for the Drude and the plasma 
model is shown as a function of the temperature as solid line for 
$\lambda_P/L=0.136$ and $\sigma_0=\sigma_\text{Au}$. The dashed line represents the 
negative thermal correction for the Drude model while the dotted line corresponds 
to the thermal correction for the plasma model.  A, B, and C indicate three 
different regimes discussed in the text.}
\label{fig:deltaeta} 
\end{figure}

The implications of the behavior of $\Delta\eta_F^T$ can be understood with the
help of Fig.~\ref{fig:deltaeta} where the solid line representing
$-\Delta\eta_F^T$ displays a smooth crossover between
$-\eta_F^T(\text{plasma})$ (dotted line) at high temperatures and
$\eta_F^T(\text{Drude})$ (dashed line) at low temperatures. Here, three
different regimes can be distinguished which are indicated in
Fig.~\ref{fig:deltaeta} by the letters A--C. Regime A corresponds to high
temperatures where to leading order in the temperature
$-\eta_F^T(\text{plasma})$ agrees with the high-temperature behavior
(\ref{eq:delta_eta_f_t0}). As the thermal corrections within the plasma model
turn to their low-temperature behavior described by (\ref{eq:lowTplasmaTE}) and
thus decrease very fast with decreasing temperature, one reaches the regime B.
Here, $\Delta\eta_F^T$ determines the thermal corrections of the Drude model
which for sufficiently small relaxation frequency $\gamma$ are still given by
the right-hand side of (\ref{eq:delta_eta_f_t0}). 

Deviations from the high-temperature behavior occur at small temperatures 
which decrease with decreasing $\gamma$. It is found in the appendix 
that there $\Delta\eta_F^T\sim T^{5/2}$ with the prefactors given in
(\ref{eq:lowTdrudeTE}). As the corresponding exponent for the plasma model is
larger, it follows that also in regime C $\Delta\eta_F^T$ agrees with the
thermal corrections for the TE mode within the Drude model. We remark that the
scenario described here holds for sufficiently small values of $\gamma$. If 
the relaxation frequency is significantly larger than the plasma frequency
$\omega_P$, an additional regime can appear as was the case in 
Fig.~\ref{fig:force_thermal} for the TM mode. In an intermediate temperature
regime, $\Delta\eta_F^T$ then follows the low-temperature behavior 
(\ref{eq:lowTplasmaTE}) of the plasma model. 

In this section, we have identified a quantity, $\Delta\eta_F^T$, which depends
on temperature only through the dimensionless quantity $\hbar\beta\gamma$.
Although this quantity describes the difference between two different models,
the Drude and the plasma models, it nevertheless directly determines $\eta_F^T$
for the Drude model at low temperatures where $\eta_F^T$ for the plasma model
in view of the strong temperature dependence (\ref{eq:lowTplasmaTE}) becomes
negligibly small. In the regions B and C of Fig.~\ref{fig:deltaeta}, the
temperature dependence of $\eta_F^T$ for the Drude model is therefore analogous 
to that of thermodynamic quantities like the entropy of a free Brownian
particle. As the relaxation frequency $\gamma$ decreases, the region B extends
increasingly farther down to low temperatures. Nevertheless, below a
temperature of the order of $\hbar\gamma/k_B$ a region C is always reached
where the low-temperature behavior (\ref{eq:lowTdrudeTE}) applies. For any
nonvanishing relaxation frequency, the situation is therefore different from
the case $\gamma=0$. There, $\Delta\eta_F^T$ simply vanishes because setting
$\gamma$ to zero in the Drude model will turn it into the plasma model.

\section{Contribution of the transverse electric mode to the entropy}
\label{sec:entropy}

We now explore how the findings in the previous section translate to the
free energy and in particular the entropy which is in the focus of the
discussion about a possible violation of the third law of thermodynamics.

The entropy can be obtained by means of the standard thermodynamic relation
\begin{equation}
S = -\frac{\partial\mathcal{F}}{\partial T}
\end{equation}
where $\mathcal{F}$ is the free energy 
\begin{equation}
\begin{aligned}
\label{eq:freeenergy}
\mathcal{F} &= \frac{\hbar A}{2\pi^2}\sum_{n=0}^\infty\strut'\int_0^\infty
d\xi\cos(n\hbar\beta\xi)\\
&\quad\times\int_{\xi/c}^\infty d\kappa\,\kappa\log\left[1-r^2(i\xi,i\kappa)
\exp(-2\kappa L)\right]\,.
\end{aligned}
\end{equation}
A Matsubara sum corresponding to (\ref{eq:f_lifshitz}) can be derived by
means of a Poisson resummation. The Casimir force and the free energy are related 
by $F=-\partial\mathcal{F}/\partial L$. Comparing the structure of 
(\ref{eq:f_resummed}) and (\ref{eq:freeenergy}), it is clear that the reasoning
of Sec.~\ref{sec:temode} can readily be applied to the free energy. In particular,
the difference between the free energies in the Drude and plasma model is also of 
the form $\Delta\mathcal{F}=\gamma g_E(\hbar\beta\gamma)$. As for the force, this
difference equals the result of the Drude model at low temperatures and thus the
scaling can readily be verified for the leading low-temperature terms in the Drude 
model 
\begin{equation}
\label{eq:lowTfreeenergy}
\begin{aligned}
\eta_E^T &= -\frac{15}{\pi^2}[2\log(2)-1]\frac{L\sigma_0}{c}\left(\frac{k_BTL}{\hbar c}\right)^2\\
&\quad-\frac{45}{2\pi^4}(2\pi)^{1/2}\zeta\left(\frac{5}{2}\right)\left(\frac{L\sigma_0}{c}\right)^{3/2}
\left(\frac{k_BTL}{\hbar c}\right)^{5/2}+\ldots
\end{aligned}
\end{equation}
which agrees with results derived earlier \cite{brevi04,hoye07} on the basis of the 
Lifshitz formula for the free energy.
In analogy to the force, we have divided in $\eta_E^T$ the free energy at temperature $T$
by the free energy 
\begin{equation}
\mathcal{F}_\text{Cas} = -\frac{\pi^2\hbar c}{720L^3}
\end{equation}
at zero temperature for ideal mirrors and accounting for both polarizations. Without
the normalization, the first term in (\ref{eq:lowTfreeenergy}) is independent of the 
mirror separation $L$ and thus does not contribute to the Casimir force. Therefore, 
it is the second term which corresponds to the leading low-temperature behavior 
(\ref{eq:lowTdrudeTE}) of the TE mode within the Drude model.

For the difference $\Delta S_\text{TE}$ of entropies in the Drude and plasma
model one finds after taking the derivative of the difference
$\Delta\mathcal{F}_\text{TE}$ of free energies with respect to temperature that
it can be written in the form $\Delta
S_\text{TE}/k_B=(\hbar\beta\gamma)^2g_E{}'(\hbar\beta\gamma)/\hbar$. Here the
prime indicates a derivative with respect to the argument of the function.  As
before for the force, we see that the limit $\gamma\to0$ leads into the
high-temperature regime. For any finite $\gamma$ there is however always a
low-temperature regime determined by (\ref{eq:lowTfreeenergy}) which ensures
that the entropy goes to zero as temperature goes to zero. 

\begin{figure}
\includegraphics[width=\columnwidth]{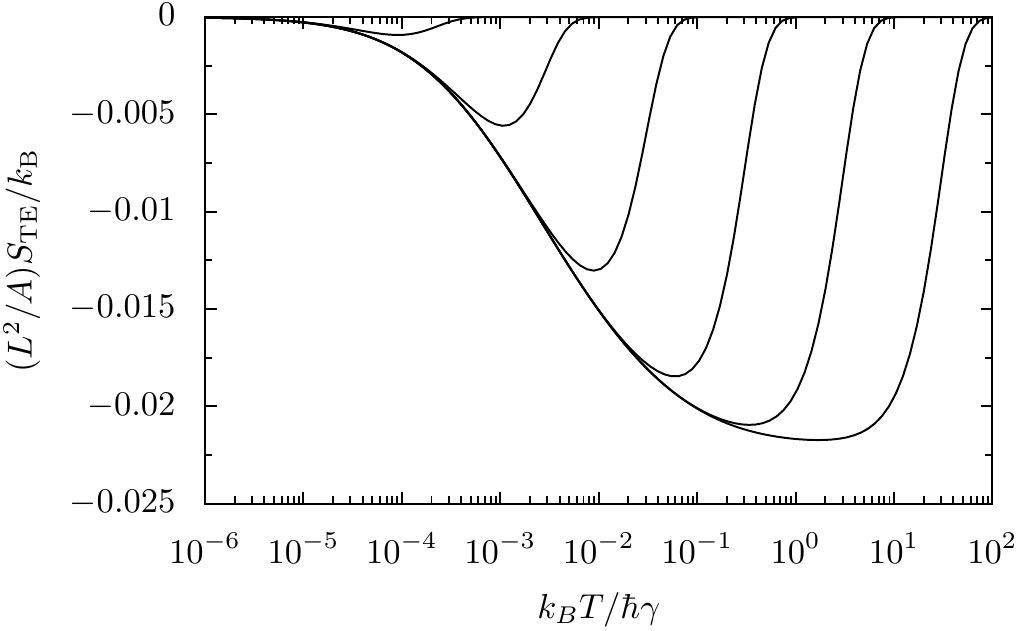}
\caption{The contribution of the TE modes to the entropy is shown as a function 
of the temperature for the Drude model with $\lambda_P/L=0.136$ and
$L\gamma/c=0.01,0.1,1,10,100,1000$ from bottom to top.}
\label{fig:entropy1} 
\end{figure}

\begin{figure}
\includegraphics[width=\columnwidth]{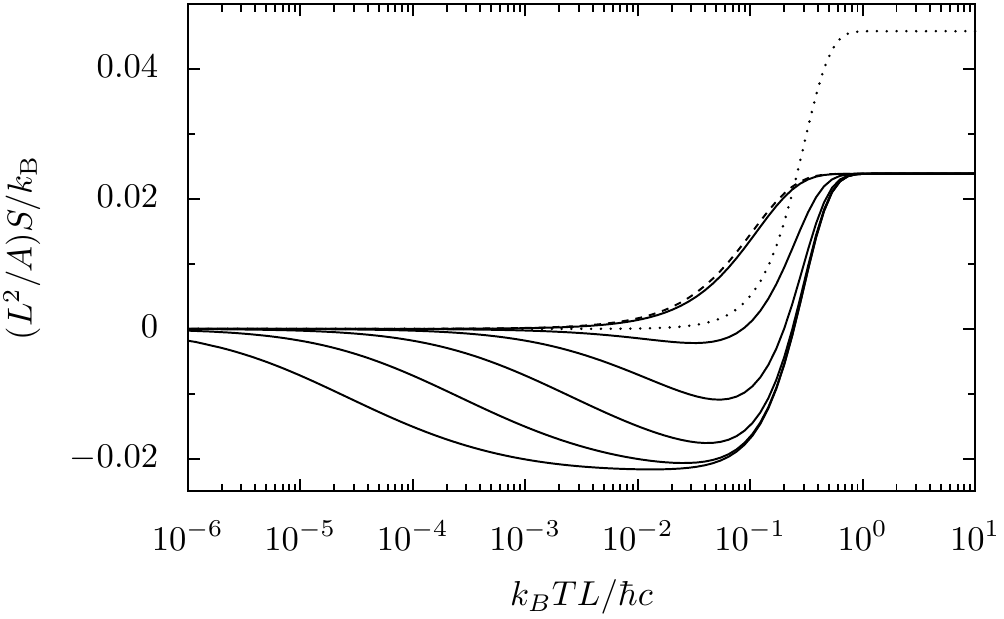}
\caption{The total entropy is shown as a function of the temperature for the 
Drude model with $\lambda_P/L=0.136$ and $L\gamma/c=0.01,0.1,1,10,100,1000$ 
(solid line, from bottom to top). The dashed line indicates the contribution of
the TM mode for $L\gamma/c=1000$ and the dotted line corresponds to the
entropy obtained for the plasma model. In contrast to Fig.~\ref{fig:entropy1}, 
temperature is taken with respect to the mirror separation $L$.}
\label{fig:entropy2} 
\end{figure}

The temperature dependence of the entropy is illustrated in
Figs.~\ref{fig:entropy1} and \ref{fig:entropy2}. Fig.~\ref{fig:entropy1} shows the
contribution of the TE mode to the entropy for the Drude model with $\lambda_P/L=
0.136$ and $L\gamma/c=0.01,0.1,1,10,100,1000$ from the lowest to the uppermost
curve. These curves demonstrate that the dependence of $\Delta S_\text{TE}$ on the 
dimensionless temperature $\hbar\beta\gamma$ indeed applies to the contribution 
of the TE mode to the entropy in the Drude model for low temperatures where the 
entropy within the plasma model is negligible.

In Fig.~\ref{fig:entropy2} the solid lines show the total entropy for the same
parameters as used in Fig.~\ref{fig:entropy1}. In addition, the dashed line represents
the contribution of the TM mode for $L\gamma/c=1000$ and the dotted line gives
the total entropy for the plasma model with $\lambda_P=0.136$. For small relaxation
frequencies $\gamma$, the temperature dependence of the entropy displays the same
qualitative behavior as found for the specific case of copper plates \cite{bostr04}.
In particular, a temperature regime exists where the entropy becomes negative.
In any case, however, the entropy will go to zero as temperature goes to zero, in
accordance with Nernst's theorem. Furthermore, it can be seen that the total entropy
remains positive if $\gamma$ is sufficiently large. This is the case for 
$L\gamma/c=1000$, \textit{i.e.}\ the uppermost curve in Fig.~\ref{fig:entropy2}.
The comparison with the corresponding contribution from the TM mode depicted as dashed
line shows that the contribution of the TE mode is nevertheless negative. 

With increasing relaxation frequency or, equivalently, decreasing conductivity,
the electromagnetic field couples less strongly to the electrons in the mirrors
and in the limit of vanishing conductivity the electromagnetic field could be
considered as an isolated system. In this situation, the entropy has to remain
positive as in fact it does. If, on the other hand, the coupling is strong, the
electromagnetic field represents only a subsystem whose entropy can well become
negative as has already been pointed out in Ref.~\onlinecite{hoye03}. 

Letting $\gamma$ formally go to zero in the low-temperature expression for the
entropy would in principle lead to a nonvanishing negative entropy at zero
temperature, but then we would have to start with the plasma model from the
very beginning where it is generally agreed that Nernst's theorem holds. The
noncontinuous transition from the Drude model to the plasma model thus makes it
possible that in the first case for nonvanishing values of $\gamma$ and in the
second case for $\gamma=0$ the entropy goes to zero in the zero-temperature
limit as it should.

\section{Conclusions}
\label{sec:conclusions}

We have considered the thermodynamics of a free damped quantum particle on the
one hand and of the electromagnetic field enclosed between two mirrors of
finite conductivity, on the other hand. If in the first case the normalization
volume is very large and in the second case the plasma frequency and the
frequency $c/L$ associated with the mirror separation $L$ are large compared to
the relaxation frequency of the Drude-type mirrors, the temperature dependence
in both cases is dominated by the interplay of two energy scales: the thermal
energy $k_BT$ and the energy $\hbar\gamma$ associated with the damping or the
relaxation present in a Drude metal. 

In such a situation, the limits of zero temperature and zero damping do not
commute.  For any finite temperature, the limit of small damping or relaxation
will restore the classical high-temperature behavior. This behavior, if
continued down to zero temperature, could potentially violate requirements
imposed by thermodynamics. However, for any nonvanishing damping, there exists
a quantum regime at low temperatures, which regularizes the approach to zero
temperature so that no problems with Nernst's theorem arise. For
room-temperature measurements on the Casimir effect with metallic mirrors made
of gold, the ratio $k_BT/\hbar\gamma$ is close to one, placing these
experiments into the transition region between the classical and the quantum
regime.

The Casimir effect with its strong coupling between the electromagnetic field
modes and the metallic mirrors is a prime example where the coupling between
system and environment is far from negligible. In such cases, the coupling will
affect the thermodynamic properties in the quantum regime and even lead to
negative values for the entropy or the specific heat at low temperatures. Such
negative values result from the restriction to a subsystem and are therefore
not in contradiction with thermodynamic principles. The appearance of a
negative entropy or specific heat is not specific to the free Brownian quantum
particle \cite{hangg08} and the Casimir effect \cite{hoye03} but is also known
for example in condensed matter physics \cite{flore04,zitko08}. 

While there exists a direct analogy in the low-temperature behavior of the free
Brownian particle and the Casimir effect, there is an apparent difference at
high temperatures.  In the first case, the limit of vanishing damping leads to
the free particle while in the second case, even in the limit of vanishing
relaxation frequency, the TE mode contributes within the plasma model while it
does not for the Drude model. This is a consequence of the fact that the
quantity which enters the analogy for the Casimir effect is not the Casimir
force itself but the difference of the Casimir forces in the Drude and the
plasma model. It is this difference which in the classical limit becomes
independent of the relaxation frequency and thus within our analogy accounts
for the suppression of the Casimir force by a factor of two at high
temperatures.

\begin{acknowledgments}
We have benefitted from useful discussions with Maguelonne Chevallier, Cyriaque 
Genet, and Carsten Henkel. GLI acknowledges financial support by the European 
Science Foundation (ESF) within the activity `New Trends and Applications of 
the Casimir Effect' (\url{www.casimir-network.com}) as well as by the ENS Paris 
during a stay at the Laboratoire Kastler Brossel.
\end{acknowledgments}

\appendix
\section{Low-temperature expressions for the Casimir force within the Drude model}
\label{sec:lowtempapprox}

In the following, we derive the low-temperature expansions (\ref{eq:lowTdrudeTE})
and (\ref{eq:lowTdrudeTM}) for the TE and TM modes, respectively, within the
Drude model. We take as our starting point the expressions (\ref{eq:f_xi})
and (\ref{eq:eta_f_t}). In view of our discussion in Sec.~\ref{sec:temode}, we introduce
a dimensionless frequency $x=\xi/\gamma$. As we are interested in the low-temperature
behavior and therefore in the behavior of $\mathbb{F}(x)$ for small $x$, it is
convenient to introduce another dimensionless variable $u=(c\kappa/\omega_P)
(1+1/x)^{1/2}$. We thus bring (\ref{eq:eta_f_t}) into the form
\begin{equation}
\begin{aligned}
\eta_F^T &= \frac{240}{\pi^4}\frac{L^4\omega_P^3\gamma}{c^4}
\sum_{n=1}^\infty\int_0^\infty dx\cos(n\hbar\beta\gamma x)\left(\frac{x}{x+1}\right)^{3/2}\\
&\quad\times\int_{u_0}^\infty du\,u^2
\frac{r_p^2(u)}{\exp\left[\frac{2L\omega_P}{c}\left(\frac{x}{x+1}\right)^{1/2}u\right]-
r_p^2(u)}
\end{aligned}
\end{equation}
with 
\begin{equation}
\label{eq:u0}
u_0 = \frac{\gamma}{\omega_P}[x(x+1)]^{1/2}\,.
\end{equation}

For the TE mode, the behavior of the integrand for small $x$ is dominated by the factor
$x^{3/2}$ in front of the integral over $u$. The lower limit of integration (\ref{eq:u0})
of that integral can then be set to zero and in its integrand, the reflection coefficient 
$r_\text{TE}(u)$ varies rapidly compared to the exponential function. To leading
order, we are therefore left with
\begin{equation}
\label{eq:eta_f_t_te}
\begin{aligned}
\eta_F^T &= \frac{240}{\pi^4}\left(\frac{L\omega_P}{c}\right)^3\frac{L\gamma}{c}
\sum_{n=1}^\infty\int_0^\infty dx\cos(n\hbar\beta\gamma x)x^{3/2}\\
&\quad\times\int_0^\infty du\,u^2\frac{r_\text{TE}(u)^2}{1-r_\text{TE}(u)^2}
\end{aligned}
\end{equation}
where
\begin{equation}
r_\text{TE}(u) = \frac{(u^2+1)^{1/2}-u}{(u^2+1)^{1/2}+u}\,.
\end{equation}
The two integrals can be carried out yielding
\begin{equation}
\int_0^\infty du\,u^2\frac{r_\text{TE}(u)^2}{1-r_\text{TE}(u)^2}=\frac{1}{12}
\end{equation}
and
\begin{equation}
\label{eq:cos_x32}
\int_0^\infty dx\cos(x)x^{3/2} = -\frac{3}{8}(2\pi)^{1/2}\,.
\end{equation}
In the latter integral, an infinitely weak exponential regularization was assumed.
Inserting these two results into (\ref{eq:eta_f_t_te}) one finds the leading 
low-temperature contribution (\ref{eq:lowTdrudeTE}) of the TE mode to the Casimir 
force.

We now turn to the TM mode. In contrast to the TE mode, the absolute value of the
reflection coefficient equals one in the limit of small frequencies, in which we
are interested in order to obtain the low-temperature behavior. We separate
the closed-loop function into two contributions
\begin{equation}
f_\text{TM}(i\xi,i\kappa) = f_\text{TM}^{(1)}(i\kappa)
+ f_\text{TM}^{(2)}(i\xi,i\kappa)
\end{equation}
with
\begin{equation}
f_\text{TM}^{(1)}(i\kappa) = \frac{1}{\exp(2\kappa L)-1}
\end{equation}
and
\begin{equation}
f_\text{TM}^{(2)}(i\xi,i\kappa) = \frac{\exp(2\kappa L)[r_\text{TM}^2(i\xi,i\kappa)-1]}
{[\exp(2\kappa L)-1][\exp(2\kappa L)-r_\text{TM}^2(i\xi,i\kappa)]}\,.
\end{equation}
The first contribution is familiar from the case of ideal mirrors and leads to a 
low-temperature contribution of the form (\ref{eq:lowTplasmaTE}) which goes with $T^4$.
The second term will turn out to increase more slowly with temperature and therefore 
determines the leading-order term. For small values of the dimensionless variables $x$ and
$u$, one obtains to leading order
\begin{equation}
u^2 f_\text{TM}^{(2)} = -\left(\frac{c\gamma}{\omega_P^2L}\right)^2\frac{1}{u}\,.
\end{equation}
The leading term of the thermal correction to the factor $\eta_F$ thus becomes
\begin{equation}
\label{eq:eta_f_t_tm_approx}
\begin{aligned}
\eta_F^T &= -\frac{240}{\pi^4}\left(\frac{L\gamma}{c}\right)^3\frac{\gamma}{\omega_P}
\sum_{n=1}^\infty\int_0^\infty dx\cos(n\hbar\beta\gamma x)x^{3/2}\\
&\quad\times\int_{\gamma x/\omega_P}^\infty du\frac{1}{u}\,.
\end{aligned}
\end{equation}
The integral over $u$ yields a logarithmic contribution from the lower limit of
integration. Such a logarithmic term also appears in the evaluation of the low-temperature
behavior of the TM mode in the plasma model \cite{borda00}, even though the final result
only contains a power of temperature. In contrast, here, the logarithmic term will 
survive the integration over $x$. The ultraviolet behavior of the integral over $u$ will
be regularized if the complete closed-loop function is taken into account. From here,
a contribution to the next-to-leading term can be expected.

Together with (\ref{eq:cos_x32}), one finds for (\ref{eq:eta_f_t_tm_approx}) the
low-temperature approximation (\ref{eq:lowTdrudeTM}) with the constant
\begin{equation}
\label{eq:constant_tm}
\begin{aligned}
C &= \frac{1}{\zeta(5/2)}\sum_{n=1}^\infty\frac{\log(n)}{n^{5/2}}\\
&\quad+\frac{8}{3(2\pi)^{1/2}}\int_0^\infty dy\cos(y)y^{3/2}\log(y)
\end{aligned}
\end{equation}
which has a numerical value $C\approx 1.155$. Note, however, that we have neglected
terms which potentially contribute to this constant.


\begin{thebibliography}{99}
\bibitem{footnote1}
In the context of dissipative quantum system, this choice of the damping kernel
is usually referred to as Drude model. To avoid confusion with the Drude model
for metals, we do not employ this term here.
\bibitem{footnote2}
In order to relate (\ref{eq:lowTdrudeTE}) to the expression in the appendix A of
ref.~\onlinecite{hoye07} note that $\zeta(-3/2) = -(3/16\pi^2)\zeta(5/2)$.
\end{thebibliography}
\end{document}